# Evolutionarily conserved human targets of adenosine to inosine RNA editing

Erez Y. Levanon[1,2,*], Martina Hallegger[3], Yaron Kinar[1], Ronen Shemesh[1], Kristina Djinovic-Carugo[4], Gideon Rechavi[2], Michael F. Jantsch[3] and Eli Eisenberg[1,5]

[1]Compugen Ltd, 72 Pinchas Rosen St, Tel-Aviv 69512, Israel, [2]Department of Pediatric Hemato-Oncology, Safra Children's Hospital, Sheba Medical Center and Sackler School of Medicine, Tel Aviv University, Tel Aviv, Israel, [3]Max F. Perutz Laboratories, Department of Chromosome Biology, University of Vienna, Rennweg 14, A-1030 Vienna, Austria, [4]Max F. Perutz Laboratories, University Departments at Vienna Biocenter, Institute for Theoretical Chemistry and Molecular Structural Biology, University of Vienna, Campus Vienna Biocenter 6/1, Rennweg 95b, A-1030 Vienna, Austria and [5]School of Physics and Astronomy, Raymond and Beverly Sackler Faculty of Exact Sciences, Tel Aviv University, Tel Aviv 69978, Israel

## ABSTRACT

**A-to-I RNA editing by ADARs is a post-transcriptional mechanism for expanding the proteomic repertoire. Genetic recoding by editing was so far observed for only a few mammalian RNAs that are predominantly expressed in nervous tissues. However, as these editing targets fail to explain the broad and severe phenotypes of ADAR1 knockout mice, additional targets for editing by ADARs were always expected. Using comparative genomics and expressed sequence analysis, we identified and experimentally verified four additional candidate human substrates for ADAR-mediated editing: FLNA, BLCAP, CYFIP2 and IGFBP7. Additionally, editing of three of these substrates was verified in the mouse while two of them were validated in chicken. Interestingly, none of these substrates encodes a receptor protein but two of them are strongly expressed in the CNS and seem important for proper nervous system function. The editing pattern observed suggests that some of the affected proteins might have altered physiological properties leaving the possibility that they can be related to the phenotypes of ADAR1 knockout mice.**

## INTRODUCTION

Site-selective adenosine to inosine (A-to-I) RNA-editing is an essential post-transcriptional mechanism for expanding the proteomic repertoire (1). It is carried out by members of the double-stranded RNA-specific ADAR family predominantly acting on precursor messenger RNAs (2). As inosines in mRNA are recognized as guanosines (G) by the ribosome in the course of translation, RNA-editing can lead to the formation of an altered protein if editing leads to a codon exchange. ADAR-mediated RNA editing is essential for the development and normal life of both invertebrates and vertebrates (3–5). Additionally, altered editing patterns have been found to be associated with inflammation (6), epilepsy (7), depression (8), ALS, (9) and malignant gliomas (10). A-to-I editing affects numerous sites in the human transcriptome, but most of these are located in non-coding regions (11–15). Three families of mammalian ADAR substrates in which editing causes amino acid substitutions were found so far, all of them encode receptors that are expressed in the CNS: subunits of the glutamate receptor superfamily(16) [see review in (17,18)], the serotonin 5-HT2C-receptor (19) and the potassium channel KCNA1 (20). In all these examples, the amino acid substitutions due to editing were shown to have a major impact on protein properties (17–19,21). All of these genes are involved in neurotransmission, pointing to a central function of RNA editing in the nervous system (20). A wide range of severe phenotypic alterations following inactivation of ADAR1 and ADAR2 has been demonstrated (5,22). However, while the RNA encoding gluRB seems the primary target responsible for the phenotype of ADAR2 deficient mice, the targets resulting in the phenotype(s) of ADAR1-deficient mice remain unknown to this point. In contrast to ADAR2, ADAR1 has a critical role in several tissues of non-neuronal origin, but the substrates in these tissues remain to be identified (22). Therefore, in the past few years, several groups have been trying to identify additional editing substrates, using both experimental and bioinformatic methods (15,20,23). Here, we describe a comparative genomics approach to identify editing events leading to amino acid substitutions.

## MATERIALS AND METHODS

### Multiple alignment

Human ESTs and cDNAs were obtained from NCBI GenBank version 139 (www.ncbi.nlm.nih.gov/dbEST). The genomic sequences were taken from the human genome build 34 (www.ncbi.nlm.nih.gov/genome/guide/human). Details of our multiple alignment (MA) model can be found in Sorek *et al*. (24).

### Details and parameters of the algorithm

In order to identify conserved editing sites, we first produced exhaustive lists of potential editing sites for mouse and human. Such potential sites are found by aligning expressed sequences (ESTs and RNAs) against their respective genomes, and finding positions where the expressed nucleotide differs from the genomic one. This process has to be done with some care as most of these mismatches are due to sequencing errors or problems in the alignment, inclusion of which would result in an enormous list of useless candidates. We used the following algorithm to identify 'true' events of disagreement between the genome and the expressed sequences, which could be either single-nucleotide-polymorphisms (SNPs) or editing sites.

The algorithm is based on a probabilistic model for the various sources of mismatches. It first aligns all available expressed sequences to the genome and clusters them into genes. For each gene, it looks for columns in the MA matrix that include mismatches, and estimates the probability of the observed nucleotide distribution being caused by either sequencing and alignment errors, or SNPs and RNA editing. If the probability for the nucleotide distribution being a result of sequencing or alignment errors does not exceed the cut-off, but the probability of an SNP/editing does, the genomic position is marked as a true event (i.e. it is assumed being an SNP or editing site). We mask sequences of low alignment quality (>10% mismatches), genomic regions where the MA is of low quality (mismatches in >20% of columns), and all single-letter repetitions and consecutive mismatches of length 3 or more. The probability of a sequencing or alignment error at a certain position is estimated based on the type of the sequence (RefSeq, RNA or EST) and the quality of the MA at the genomic region (error probabilities: clean regions—RefSeq: 2e-6; RNA: 8e-5; EST: 3e-3, dirty regions—RefSeq: 5e-6, RNA: 5e-4, EST: 8e-3, polluted regions—RefSeq: 8e-4; RNA: 5e-3; EST: 5e-2). The probability cut-off against which the different model probabilities are compared is $10^{-6}$ divided by the number of supporting sequences. The prior probability of a two-alleles SNP is $10^{-4}$. Applying this algorithm to the human and mouse transcriptomes resulted in two lists of putative SNPs/editing events.

Subsequently, the sites found in the human genome were aligned against those found in the mouse genome, retaining only alignments longer than 50 nt with identity levels higher than 85%, and nucleotide mismatches occurring at identical positions within the two sequences. We also eliminated genomic sites that are duplicated in either genome, and retained only non-synonymous events in the coding sequence.

### Experimental protocols

For human sequences, total RNA and genomic DNA (gDNA) isolated simultaneously from the same tissue sample were purchased from Biochain Institute (Hayward, CA). In this work, we used samples of liver, prostate, uterus, kidney, lung normal and tumour, brain tumour (glioma), cerebellum and frontal lobe.

The total RNA underwent oligo-dT primed reverse transcription using Superscript II (Invitrogen, Carlsbad, CA) according to manufacturer's instructions. The cDNA and gDNA (at 0.1 µg/µl) were used as templates for PCR reactions. We aimed at high sequencing quality and thus amplified rather short genomic sequences (roughly 200 nt). The amplified regions chosen for validation were selected only if the fragment to be amplified maps to the genome at a single site. PCR reactions were done using Abgene ReddyMix™ kit (Takara Bio, Shiga, Japan) using the primers and annealing conditions as detailed in the following. The PCR products were run on 2% agarose gels and only if a single clear band of the correct approximate size was obtained, it was excised and sent to Hy-labs laboratories (Rehovot, Israel) for purification and direct sequencing without cloning.

For mouse and chicken sequences, poly-A RNA was isolated from brain and liver samples using Trifast (PeqLab, Germany) and poly-A selected using magnetic oligo dT beads (Dynal, Germany). Poly A RNA (1 µg) was reverse-transcribed using random hexamers as primers and RNAseH deficient M-MLV reverse transcriptase (Promega, Madison, WI). Genomic DNA from the same samples was isolated according to Ausubel *et al*. (25).

First strand cDNAs or corresponding genomic regions were amplified with suitable primers using Pfu polymerase to minimize mutation rates during amplification. Amplified fragments were A-tailed using Taq polymerase, gel purified and cloned into pGem-T easy (Promega, Madison, WI). After transformation in *E.coli*, individual plasmids were sequenced and aligned using ClustalW.

We used Sequencher 4.2 Suite (Gene Codes Corporation) for multiple-alignment of the electropherograms (see Supplementary Figure). The extent of A–I editing varies: the level of the guanosine trace is sometimes only a fraction of the adenine trace, while in some occasions the conversion from A to I is almost complete. For each gene tested, we sequenced the PCR products derived from the three tissues in which the expression was the highest. The RT–PCR and gDNA–PCR products of one of these tissues were sequenced from both ends to ensure the consistency of the resulting electropherograms.

### Molecular modeling

The three-dimensional model of human alpha-filamin repeat 22 was modelled on structures of human gamma-filamin repeat 23 (26 and manuscript in preparation) and on the repeat 4 the actin cross-linking gelation factor APB120 from *Dictyostelium discoideum* (entry codes 1ksr, 1wlh) (27–29). Repeat 5 was refused by the server due to low sequence homology, while for domain 6 only the C-terminal part was accepted and modelled.

The homology modelling servers SwissModel (30), instead of 3D-JIGSAW (31,32) were used for the purpose. Figures were prepared with Pymol [DeLano, W. L. (2002) DeLano Scientific, San Carlos, CA, USA. http://www.PyMOL.org].

## RESULTS

*In silico*, editing may be detected using the large-scale database of ESTs (33) and RNAs, currently holding over 5 million human records. Editing sites show up when a sequence is aligned with the genome: while the DNA reads A, sequencing identifies the inosine in the edited site as G. However, most such misalignments indicate sequencing errors or genomic variability due to SNPs, rather than RNA editing. In order to eliminate these instances, we use the fact that such false-positive events are not conserved between species. Furthermore, functional editing regions are found to be highly conserved, due to constraints on the dsRNA structure required for the editing (20). Thus, we searched for conserved genomic loci in which an A-to-G mismatch appears in both human and mouse at the same position. More details on the algorithm are given in Materials and Methods. Using this approach, after screening for putative sites affecting the amino acids and filtering out duplicated genes, we found four novel editing targets, the RNAs encoding BLCAP, FLNA, CYFIP2 and IGPFB7, as well as the known editing substrates encoding glutamate receptor subunits. Interestingly, neither the RNA encoding serotonin receptor HT2C nor the KCNA1 encoding RNA were identified in this screen, due to the low representation of edited sequences covering these genes in the EST dataset (see Discussion). Our algorithm is designed to find genomic sites at which the expressed nucleotide diverges from the genomic one. Such occurrences could be interpreted as either SNPs or editing, and it is therefore not surprising to find that all of the editing sites reported here are erroneously recorded as SNPs in dbSNP (dbSNP ids: BLCAP—rs11557677; FLNA—rs3179473; CYFIP2—rs3207362; IGFBP7—rs1133243 and rs11555284). All of these presumed SNPs have no evidence for genomic polymorphisms, and were included in dbSNP based on expressed data alone. One therefore has to consider the possibility of editing rather than genomic polymorphism for dbSNP records based solely on expressed data.

To experimentally validate the predicted editing sites, we sequenced matching DNA and RNA samples retrieved from the same specimen, for up to six tissues of human and mouse. Additionally, brain and liver cDNA and genomic DNA were sequenced for the chicken FLNA and CYFIP2 genes. We verified editing events in all predicted substrates in human and mouse, except for the mouse IGFBP7 gene, which we failed to amplify successfully. In addition, we verified editing of CYFIP2 and FLNA in chicken tissues (Figures 1 and 2). PCR products were either cloned followed by sequencing of individual clones (mouse and chicken), or sequenced as a population without cloning (human). When the PCR products were cloned, the occurrence of editing was detected by comparing the sequences of several clones with the genomic sequence. When PCR products were directly sequenced,

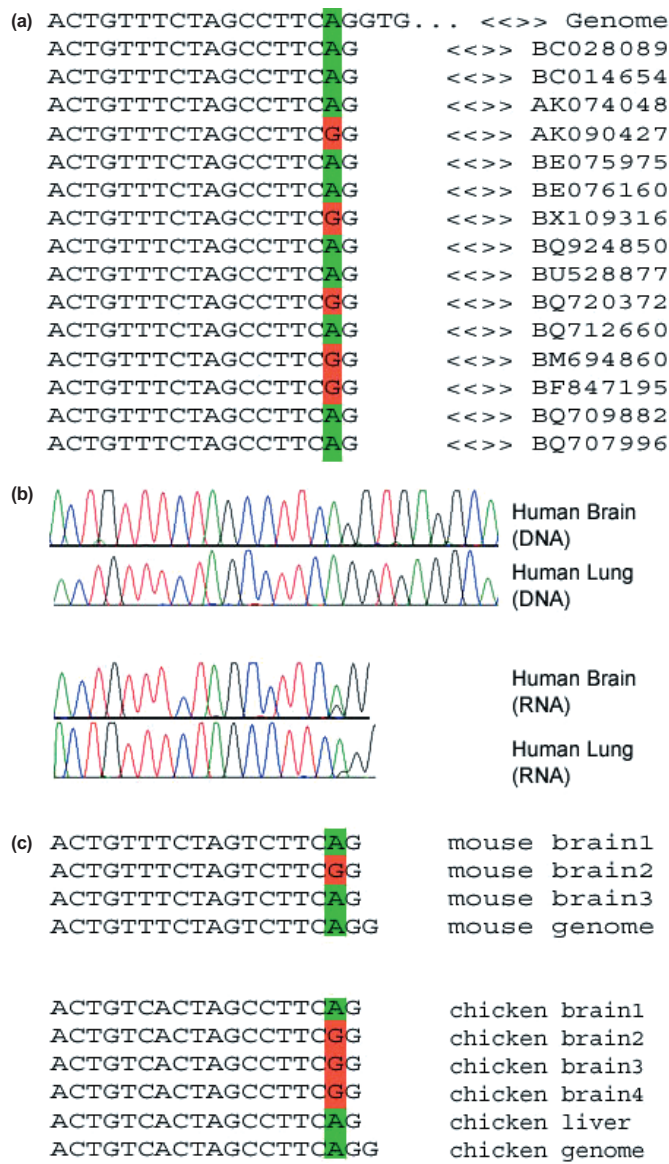

**Figure 1.** Editing in FLNA transcripts. (**a**) Some of the publicly available expressed sequences covering this gene, together with the corresponding genomic sequence. A total of 226 sequences are available for this locus, 23 of which are edited. (**b**) Results of sequencing experiments. Matching human DNA and cDNA RNA sequences for human brain and lung tissues. Editing is characterized by a trace of guanosine in the cDNA RNA sequence, where the DNA sequence exhibits only adenosine signals. Sequencing data for more tissues are available as Supplementary Material. Note the variety of tissues showing editing and the variance in the relative intensity of the edited guanosine signal. (**c**) Sequences of individually cloned fragments from matching DNA and RNA of mouse brain tissues and chicken brain and liver tissues. Only part of the data is shown. A total of 20 mouse brain cDNA clones, 10 chicken brain and 9 chicken liver cDNA clones were sequenced, out of which four, seven and one sequence showed editing events, respectively. Similar results for the other two substrates are provided as Supplementary Material.

editing was determined by the presence of an unambiguous trace of guanosine in positions for which the genomic DNA clearly indicated the presence of an adenosine. In some cases, where individual clones were sequenced, additional A/G mismatches were found adjacent to the predicted site (see Supplementary Material). These additional sites

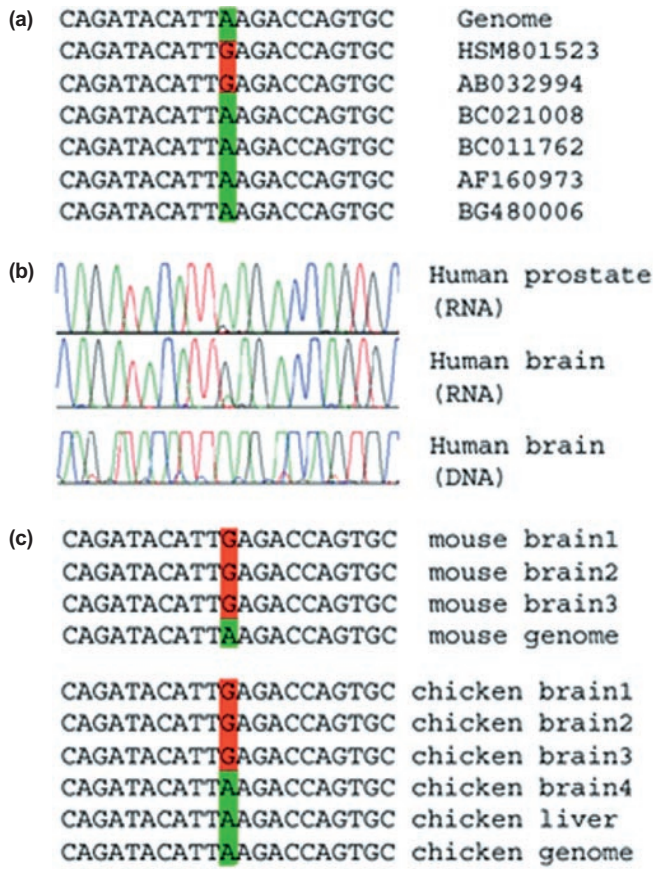

**Figure 2.** Editing in CYFIP2 transcripts. (**a**) Some of the publicly available expressed sequences covering this gene, together with the corresponding genomic sequence. A total of 23 sequences are available for this locus, two of which are edited. Both edited sequences originate from brain tissues. (**b**) Results of sequencing experiments. Matching human DNA and cDNA sequences for human brain and prostate cDNA. As in Figure 1, editing is characterized by a trace of guanosine in the cDNA RNA sequence, where the DNA sequence exhibits only adenosine signals. Sequencing data for more tissues are available in Supplementary Figure. In the brain, the editing signal surpasses the original adenosine signal, but in other tissues it is very weak. (**c**) Sequences of individually cloned fragments from matching DNA and RNA of mouse brain tissues and chicken brain and liver tissues. Only part of the data is shown. A total of eight mouse brain cDNA clones were sequenced and all of them were edited. Nine chicken brain cDNA clones were sequenced, out of which four were edited. In contrast, none of the eight chicken liver cDNA clones was edited. These results suggest that editing of this site might be brain specific, in agreement with the data for human tissues presented in the previous panel. Similar results for the other two substrates are provided as Supplementary Material.

occurred at a low frequency and were therefore missed by the search algorithm.

The full-length BLCAP (bladder cancer associated protein) cDNA contains a complete open reading frame (ORF) encoding a protein composed of 87 amino acids. Comparison of mouse and human BLCAP genomic loci revealed an intronless organization of the coding region in both species as well as a highly conserved structure having 91% and 100% identity at the DNA (coding region) and protein levels. The function of this differentially expressed protein is not yet known, but it is expressed mainly in brain tissues and B cells (34) and appears to be down-regulated during bladder cancer progression (35). We identified an editing site within the BLCAP coding sequence, located at chr20:36 833 001 (here and in what follows, we use UCSC coordinates on the July 2003 build of the human genome), inducing a Y→C substitution at the second amino acid of the final protein. There is a highly conserved region within the intron, ~500 bp upstream of the editing site, which potentially pairs with the editing region to form an almost perfect, 48 bp long, dsRNA hairpin structure (Figure 3). Notably, our experimental results show evidence for an additional U→C editing site at chr20:36 832 971, resulting in an L→P substitution (data not shown).

The FLNA (filamin A) protein is a 280 kD (2647 amino acid) protein that cross-links actin filaments into orthogonal networks in the cortical cytoplasm (36) and participates in the anchoring of membrane proteins with the actin cytoskeleton (37). The resulting remodelling of the cytoskeleton is central to the modulation of cell shape and cell migration. We identified one editing site within the FLNA transcript (chrX:152 047 854) resulting in a Q→R substitution at amino acid 2341 in the human and mouse proteins, and 2283 in the chicken homologue. The human editing region is predicted to form a 32 bp long dsRNA structure with a conserved region within the intron ~200 bp downstream to the editing site. The edited amino acid lies within the 22nd Immunoglobulin-like domain of the protein, which has been shown to be important for interaction with integrin beta (38). The same region binds to the small GTPase Rac1 (39) involved in cytoskeletal reorganization, which is also known to interact with the *Drosophila* homologue of CYFIP2, another target identified in our screen (40).

The biological consequence of the introduced Q to R substitution in the FLNA protein remains to be determined. Therefore, to get a better understanding of how the introduced amino acid exchange might alter protein function, we attempted to model the amino acid substitution on available crystal structures of two related domains. One of these is the 4th repeat of ABP120 from *Dictyostelium*, the other is repeat 23 of gamma-filamin, a close homologue of FLNA. Both available structures adopt an immunoglobulin-like fold. The amino acid sequence alignment of human alpha-filamin repeat 22, where the edited amino acid is located, with human gamma-filamin repeat 23 shows 41% sequence identity; the sequence alignment with the Dictyostelium domain 4 gives ~22% sequence identity, and 20% with domains 5 and 6, which were not used for modelling studies.

In repeat 23 of gamma-filamin as well as in domain 4 of ABP120, the position corresponding to the edited Gln2341 in FLNA is occupied by a Gly residue. In both available structures, the corresponding Gly resides at the beginning of a long loop region that connects the first and the second beta-strand. At its N-terminus, this loop locally adopts a beta-turn structure in gamma-filamin. Substitution of the polar Gln with the positively charged Arg in the edited protein requires some local rearrangements in order to accommodate the long polar sidechain. Despite higher sequence identity between repeat 23 of gamma-filamin and FLNA, the local conformation of the loop hosting the edited site was less suitable as a template for modelling: the side chain of Arg introduced at position 2341 replacing the resident Gln points towards the core of the protein, leading to a steric clash. On the other hand, in domain 4 of ABP120, the edited Arg residue can be accommodated easily without major rearrangements (see Figure 4).

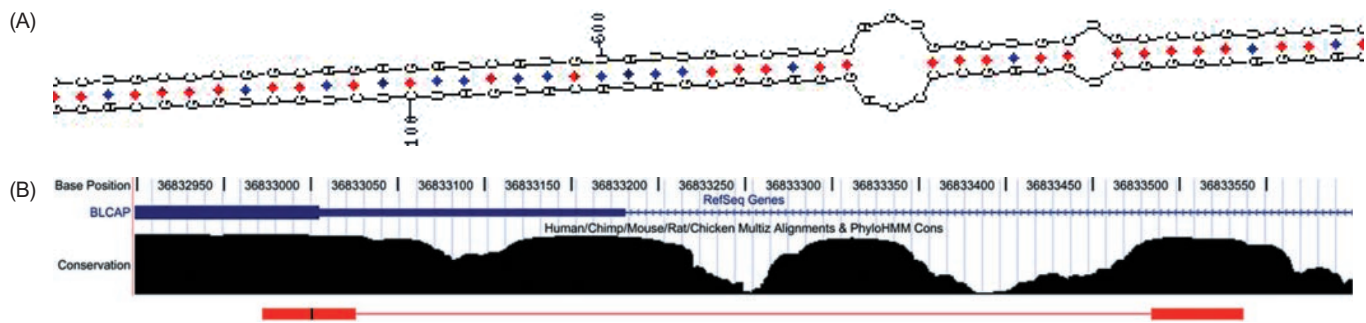

**Figure 3.** Hairpin structure in BLCAP transcripts. (**a**) The predicted secondary structure for the BLCAP substrate, based on lowest free-energy predictions using the program MFOLD (50) (www.ibc.wustl.edu/~zuker/rna/). The editing site is at position 601, where the codon UAU(Y) is edited into UGU(C). Structures for the other substrates are given in Supplementary Material. (**b**) Conservation levels at the editing genomic locus. The two red bars at the bottom mark the editing region and the intronic sequence almost perfectly pairing with it to form the hairpin structure shown in (a). The editing site is marked in black within the left red bar. The high conservation level of the intronic sequence, suggesting a functional importance, supports its identification as necessary for the editing process.

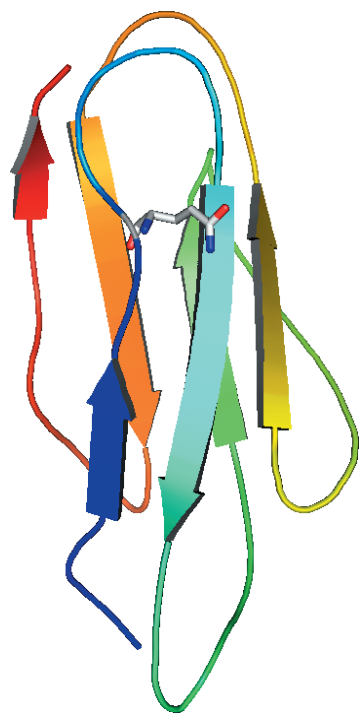

**Figure 4.** Ribbon representation of the molecular model of alpha-filamin repeat 22 generated using the Ig-like repeat from ABP120 as the template. The edited residue Gln is highlighted.

Obviously, the electrostatic potential of the domain is altered by the changed amino acid, possibly resulting in modifications of the interactions with binding partners.

The CYFIP2 (cytoplasmic FMR1 interacting protein 2) transcript encodes a protein of 1253 amino acids. CYFIP2 is a member of a highly conserved protein family found in both invertebrates and vertebrates. Human CYFIP2 shares ~99% sequence identity with its mouse orthologues (41). It is expressed mainly in brain tissues, white blood cells and the kidney (34). We identified one editing site within the CYFIP2 transcript (chr5:156 717 703) resulting in a K→E substitution at amino acid 320 in both the human and mouse proteins. Editing was also observed at the corresponding predicted position in the chicken cDNA. We note that while a strong editing signal was observed for human cerebellar cDNA, only a residual signal was observed in human lung, prostate and uterus tissues. This pattern is in agreement with the results in mouse and chicken: all eight mouse brain clones and four out of nine chicken brain clones were edited, while none of the eight chicken liver clones was edited. CYFIP2 is a p53-inducible protein (42), thus possibly a pro-apoptotic gene. Interestingly, ADAR1 knockout mice show elevated apoptosis in most tissues therefore providing a possible link between the phenotype of these mice and a potential pro-apoptotic editing target (23). No obvious dsRNA structure in the CYFIP2 pre-mRNA including the editing region could be identified, except for a weak, local pairing.

The IGFBP7 (insulin-like growth factor binding protein 7) transcript encodes a protein of 282 amino acids in length, and is expressed in a wide range of tissues (34). IGFBP7 is a member of a family of soluble proteins that bind insulin-like growth factors (IGFs) with high affinity. Their principal functions are to regulate IGF availability in body fluids and tissues and to modulate IGF binding to its receptors (43). We identified two editing sites within the IGFBP7 transcript (chr4:57 891 828 and chr4:57 891 776) resulting in R→G and K→R substitutions at amino acids 78 and 95, respectively (although we had failed to amplify the genomic region of IGFBP7, editing signals can be seen in the RNA sequences; see Supplementary Figure). The editing region seemingly pairs with a region within the coding sequence, 200 bp upstream of the editing site, to form a 140 bp long dsRNA structure. In addition, the editing site overlaps with an intron of an antisense transcript BC039519, pairing with this RNA could also trigger editing by ADARs (44).

The two edited sites in IGFBP7 map to the insulin growth factor binding (IB) domain of IGFBP7. The structure of this modulus from the homologous protein IGFBP5 was determined in solution as an isolated molecule (45) as well as in a complex with the insulin-like growth factor I (IGF-I) (46). In this case, the proteolytically stable mini-IGFBP5 construct, which retains high affinity binding to IGF-I, was used. The amino acid sequence identity between the IB domains of IGFBP7 and IGFBP5 is 29%. We carried out a structural analysis of the edited sites using the available three-dimensional model of the mini-IGFBP5 in complex with IGF-I.

The two amino acids corresponding to the edited residues are not directly involved in binding to IGF-I, but closely flank the regions involved in binding. The first edited site at position 78 (51 in IGFBP5) is close to the position Val-49 (IGFBP5 numbering) which is involved in an important hydrophobic interaction with Phe-16 of IGF-I. Val-49 is located in a loop region. In native IGFBP7, position 78 (51 in IGFBP5) hosts an Arg, the side-chain of which does not point towards the complex interface. A non-conservative substitution/mutation to glycine at this position could introduce additional flexibility and consequent change of the loop conformation, therefore disturbing the hydrophobic interaction that stabilizes the complex.

The second edited site 95 (68 in IGFBP5) hosts Lys in IGFBP5 as well as in IGFBP7, and is solvent exposed. It makes a couple of non-specific interactions via the aliphatic part of the side-chain with Glu-38 of IGF-I. In the edited molecule, this position is occupied by Arg, the long side-chain of which can maintain these weak interactions.

Notably, the editing site in the FLNA transcript is located two nucleotides upstream of a splicing site, resembling the R/G editing site in the gluRB transcript. In addition, seven of the eight nucleotides around the editing site are identical in the two substrates. This might suggest that editing and splicing in both the FLNA transcript and the R/G site in gluRB might be regulated similarly. The proximity of the editing site in the glutamate receptor to the splicing site has led to speculations on a possible link between editing and splicing. Indeed, *in vitro* splicing reactions have shown that the presence of ADAR2 inhibits splicing of the R/G site in gluRB RNA (47). Interestingly, analysis of the available EST data shows a positive correlation between editing of the last codon in the exon of FLNA and aberrant retention of the following intron, suggesting a link between editing and splicing.

## DISCUSSION

Here, we have provided evidence for ADAR-mediated RNA-editing in four novel coding substrates. While this number may seem low, we believe that many coding substrates may have been missed by our algorithm for several reasons: Editing typically happens in only a fraction of the RNAs transcribed from a given gene. Since the coverage of expressed sequences is scarce for many genes, editing sites might be missed in these cases. For example, we failed to detect the editing of the serotonin receptor, which is supported by only one edited human mRNA sequence (accession code AF208053). Similarly, no edited human KCNA1 sequence is found in the database. Also, only some of the editing sites were detected in GluR transcripts, depending on their EST coverage and the level of editing in human and mouse. In addition, the search parameters used here were rather strict, resulting in a small but accurate set of candidates. Therefore, improvements of the algorithm, more liberal parameters, and the continuous growth of the public EST databases will almost certainly lead to the identification of additional candidate editing sites.

The human proteins affected by ADAR editing found so far are all neuronally expressed receptors. In addition, ADAR2 knockout mice, as well as *adr* knockout flies show behavioural phenotypes (4). Therefore, it was hypothesized that A-to-I RNA editing has a pivotal role in nervous system functions (4). Notably, while all four novel substrates presented here do not encode for receptors, at least two of them have functions in the CNS. CYFIP2 interacts with the Fragile-X mental retardation protein (41), as well as with the FMRP-related proteins FXR1P and FXR2P, and is present in synaptosomal extracts (41). The *Drosophila* homologue has also been shown to be required for normal axonal growth and synapse formation (40,48). In addition, our experimental results suggest that the editing of CYFIP2 is brain specific. Most notably, FLNA binds a plethora of transmembrane receptors and ion channels (37). Mutations in FLNA are associated with periventricular nodular heterotopia, a disorder of neuronal migration characterized by nodules of heterotopic neurons abutting the lateral cerebral ventricles (49). The neurological features of this condition range from asymptomatic state to severe drug-resistant epilepsy. However, while CYFIP2 seems to be edited mainly in the brain, editing of FLNA, as well as BLCAP and IGFBP7 is observed in a broad range of tissues, in accordance with the expression spectrum of ADAR1. Thus, while this work provides additional support for the importance of RNA editing for CNS functions, some of the novel targets identified here may be involved in different physiological processes and could thus represent ADAR1 targets (22).

## SUPPLEMENTARY MATERIAL

Supplementary Material is available at NAR Online.

## ACKNOWLEDGEMENTS

We thank Rotem Sorek for his useful suggestions and comments on this work, and Shulamit Michaeli for helpful discussion. The work of E.Y.L. was done in partial fulfilment of the requirements for a PhD degree from the Sackler Faculty of Medicine, Tel Aviv University, Israel. Part of this work was supported by the Austrian Science Foundation grant SFB1706 to M.F.J. Funding to pay the Open Access publication charges for this article was provided by Compugen Ltd.